\documentclass[aps,prl,twocolumn,groupedaddress,superscriptaddress,amsfonts,amssymb,amsmath,citeautoscript,longbibliography,a4paper]{revtex4-2}

\usepackage{graphicx}   
\usepackage{import}                         
\usepackage{amsmath} 
\usepackage{bm}
\usepackage{amssymb}
\usepackage{xspace}
\usepackage{layouts}
\usepackage{transparent}
\usepackage{multirow}
\usepackage{mdframed}
\usepackage{color}
\usepackage{xcolor,soul}
\usepackage{bm}
\usepackage{dsfont}
\usepackage{slashed}
\usepackage{braket}
\usepackage{svg}
\usepackage[export]{adjustbox}
\usepackage{hyperref}
\hypersetup{colorlinks,
	linkcolor={blue!75!black!80!yellow},
	citecolor={blue!75!black!80!yellow},
	urlcolor={blue!75!black!80!yellow}
}

\usepackage[capitalize,nameinlink]{cleveref}
\crefname{subequations}{Eqs.}{Eqs.} 
\Crefname{subequations}{Eqs.}{Eqs.}
\crefformat{subequations}{#2Eqs.~(#1)#3}
\Crefformat{subequations}{#2Eqs.~(#1)#3}
\crefname{page}{p.}{p.} 
\usepackage{natbib}

\usepackage{textcomp} 

\renewcommand{\paragraph}[1]{\vskip 1ex\noindent\textbf{#1.}~}
\usepackage[utf8]{inputenc}

\begin{document}
\rmfamily

\title{
Alkali Intercalation of Moire Heterostructures for Low-Loss Plasmonics
}

\newcommand{\mitaffil}{Department of Physics, Massachusetts Institute of Technology, Cambridge, MA 02139, USA}
\newcommand{\rpiaffil}{Materials Science and Engineering, Rensselaer Polytechnic Institute, Troy, NY 12180, USA}
\newcommand{\cornellaffil}{School of Applied and Engineering Physics, Cornell University, Ithaca, NY 14853, USA}
\newcommand{\harvardaffileng}{School of Engineering and Applied Sciences, Harvard University, Cambridge, MA  02138, USA}
\newcommand{\harvardaffil}{ Department of Physics, Harvard University, Cambridge, MA 02138, USA}
\newcommand{\isnaffil}{Institute for Soldier Nanotechnologies, Massachusetts Institute of Technology, Cambridge, MA 02139, USA}

\author{Ali~Ghorashi}
\email{aligho@mit.edu}
\affiliation{\mitaffil}

\author{Nicholas Rivera}
\affiliation{\mitaffil}
\affiliation{\harvardaffil}
\affiliation{\cornellaffil}

\author{Ravishankar Sundararaman}
\affiliation{\rpiaffil}

\author{Efthimios Kaxiras}
\affiliation{\harvardaffil}
\affiliation{\harvardaffileng}

\author{John Joannopoulos}
\affiliation{\mitaffil}
\affiliation{\isnaffil}

\author{Marin~Solja\v{c}i\'c}
\affiliation{\mitaffil}

\begin{abstract}
    Two-dimensional metals generically support gapless plasmons with wavelengths well below the wavelength of free-space radiation at the same frequency. Typically, however, this substantial confinement of electromagnetic energy is associated with commensurately high losses, and mitigating such losses may only be achieved through judicious band structure engineering near the Fermi level. In a clean system, an isolated, moderately flat, band at the Fermi level with sufficiently high carrier density can support a plasmon that is immune to propagation losses up to some order in the electron-phonon interaction. However, proposed materials that satisfy these criteria have been ferromagnetic, structurally unstable, or otherwise difficult to fabricate. Here, we propose a class of band structure engineered materials that evade these typical pitfalls---Moire heterostructures of hexagonal boron nitride intercalated with alkali atoms. We find that only sodium atoms engender a sufficiently isolated band with plasmons lossless at first order in the electron-phonon interaction. We calculate higher order electron-phonon losses and find that at frequencies of about $1$eV the electron-phonon decay mechanism is negligible---leading to a contribution to the decay rate of $\approx 10^{7} s^{-1}$ in a small frequency range. We next calculate losses from the electron-electron interaction and find that this is the dominant process---leading plasmons to decay to lower frequency plasmons at a rate of around $10^{14} s^{-1}$. 
\end{abstract}

\maketitle
\section{Introduction}
Plasmons are collective density modes of electron liquids mediated by the long-range part of the Coulomb interaction \cite{pines2018theory, bohm1953collective}. Crucially, they have wavelengths far below the wavelength of light in vacuum \cite{jablan2013plasmons}, making them promising for applications in fields such as photovoltaics \cite{atwater2010plasmonics}, high-harmonic generation \cite{wokaun1981surface}, and sensing \cite{yee1999surface}. Unfortunately, even in the absence of impurities, plasmons are susceptible to losses due to the electron-phonon interaction \cite{jablan2009plasmonics, stauber2008effect}. Conceptually, such losses can be understood as processes in which the energy of the collective mode is transferred to an electron-hole pair and one or more quanta of lattice oscillations. 

One way to avoid electron-phonon-assisted plasmonic loss (up to some number of emitted/absorbed phonons) is by creating a material with an isolated flat band at the Fermi level \cite{ghorashi2024highly, khurgin2010search, khurgin2012reflecting, gjerding2017band, hu2022high, beida2025correlation}. If the maximum phonon energy is $\hbar\omega_\text{ph}$ and the bandwidth of the isolated band is $W$, a plasmon with energy $\hbar\omega>\hbar\omega_\text{ph}+W$ cannot decay via an intraband process into an electron-hole pair and a single phonon. Furthermore, if the closest valence/conduction band to the Fermi level is at energy, $E_{\text{v, c}}$,  a plasmon with frequency $\hbar\omega < |E_{\text{v, c}}-E_{\text{Fermi}}|-\hbar\omega_{\text{ph}}$ cannot decay into an interband electron-hole pair and a single phonon. 

In a previous work \cite{ghorashi2024highly}, we showed that lattices of substitutional defects in hexagonal boron nitride (hBN) engender flat bands at the Fermi level and that, for appropriate levels of doping, a $3\times 3$ lattice of subsitutional carbon atoms in hBN can support a plasmon that is lossless at first order in the electron-phonon interaction. However, we also found that all lattices larger than $\sqrt{3}\times\sqrt{3}$ are ferromagnetic, because of the Stoner mechanism, and thus have to be doped to be metallic. In addition, due to the significant interband screening of monolayer hBN (as well as the reduced electronic density due to spin-splitting), we found that the maximum achievable "lossless" plasmon frequency is $\approx 0.4$eV in such systems.

\begin{figure*}
    \centering
    \includegraphics{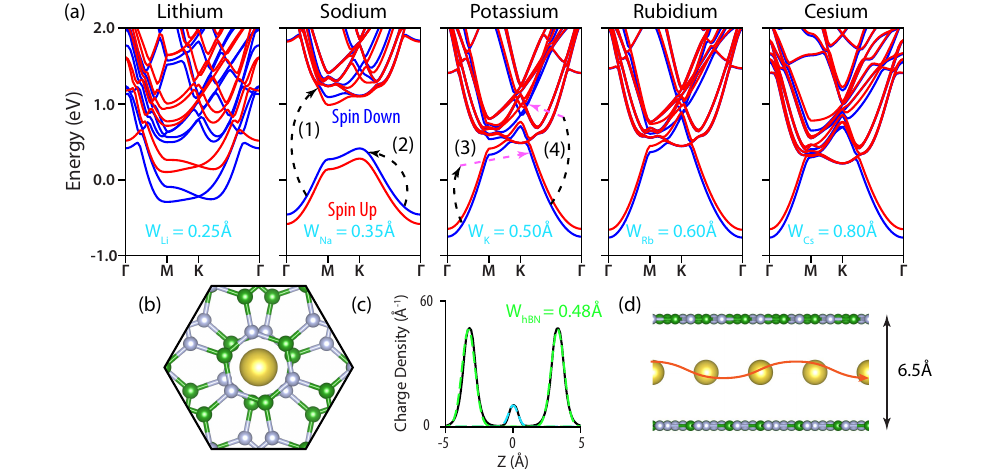}
    \caption{\textbf{$\sqrt{7}\times \sqrt{7}$ Moire physical and electronic structures}: (a) Electronic band structures of the five $\sqrt{7}\times\sqrt{7}$ systems investigated in this paper. $(1), (2)$ denote interband and intraband direct transitions, respectively. (3), (4) denote intraband and interband phonon-assisted transitions, respectively. The quantities, $\text{W}_{\text{X}}$ where X is an Alkali atom refer to the width of the intercalated layer (shown in part (c) as the middle Gaussian). (b) Top view of the physical structure. Green and white atoms are boron and nitrogen, respectively. The yellow atom is the intercalated alkali atom. (c) Z-profile of the charge density. The dashed green and blue lines are Gaussian fits. (d) Side view of the physical structure, shown for the case of intercalated sodium. The red line is a schematic of the plasmon, propagating horizontally and localized on the alkali layer.}
    \label{figure: figure 1}
\end{figure*}

Here, we explore a different class of materials---Moire heterostructures of hBN intercalated with alkali atoms (see \cref{figure: figure 1} (b)). The motivation behind exploring this class of materials is threefold. Importantly, the Moire potential provides a natural potential energy landscape to hold alkali atoms in place, making this class of materials more amenable to experimental realization than lattices of substitutional defects. In addition, the perpendicular distance between the hBN layers and the metallic alkali layer mitigates interband screening. Lastly, as we show later, sufficiently large angle Moire systems are immune to the Stoner mechanism, and thus do not need to be doped to be metallic. 

We note that non-Moire intercalated heterostructures and small-angle Moire systems have been studied in the past for their plasmonic properties \cite{papaj2020plasmonic, lewandowski2019intrinsically, lonvcaric2018strong, rukelj2020dynamical}. Our work differs from \cite{papaj2020plasmonic, lewandowski2019intrinsically} in that here we are concerned with large-angle Moire systems that potentially enable plasmons immune not just to Landau damping but to electron-phonon-assisted loss. In addition, our work differs from \cite{lonvcaric2018strong, rukelj2020dynamical} in that we are exploring systems with a natural placement for the intercalated atoms, engendered by the Moire potential. 

\begin{figure*}
    \centering
    \includegraphics{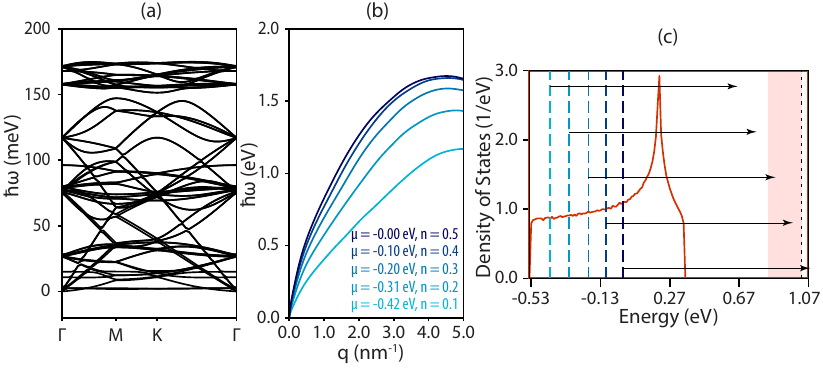}
    \caption{\textbf{Decay phase space for the sodium intercalated Moire heterostructure}. (a) Phonon dispersion indicating no dynamical instabilities and a maximum phonon frequency of $\approx0.18$ eV. (b) Plasmon dispersion for five values of hole doping. $\mu=0$ corresponds to no doping (a half filled band at the Fermi level). (c) Electronic density of states (per spin per unit cell) with the five Fermi levels of interest superimposed as dashed lines. Solid arrows indicate a plasmon of the minimum frequency to evade intraband electron-phonon assisted losses. Arrows that enter the shaded pink region are susceptible to interband electron-phonon assisted losses. Black vertical dashed line indicates the frequency at which the next conduction band starts.}
    \label{figure: figure 2}
\end{figure*}

We explored ten Moire heterostructures in total, corresponding to five alkali elements and two  Moire angles. The chosen Moire angles corresponded to unit cells that were $\sqrt{7}\times\sqrt{7}$ and $\sqrt{13}\times\sqrt{13}$ the size of the hBN primitive unit cell. We found that the five $\sqrt{13}\times\sqrt{13}$ heterostructures are susceptible to ferromagnetism, as shown in the Supplemental Materals (SM), and that only one of the $\sqrt{7}\times\sqrt{7}$ structures (corresponding to a sodium intercalant) has an isolated, moderately flat, band at the Fermi level (see \cref{figure: figure 1}(a)). We focus on this heterostructure for the remainder of the paper. Importantly, we find that, after structural relaxation, this system is  stable, as verified by the absence of imaginary phonon frequencies (see \cref{figure: figure 2}(a)).    

To calculate the plasmon dispersion, we used a three-layer model:
\begin{equation}
    -\frac{i|\mathbf{q}|}{2\epsilon_0 \omega}\sigma(\mathbf{q}, \omega)= \frac{e^{|\mathbf{q}|d}+\cosh(|\mathbf{q}|d)\frac{i|\mathbf{q}|}{\epsilon_0 \omega}\tilde{\sigma}}{e^{|\mathbf{q}|d}+\sinh(|\mathbf{q}|d)\frac{i|\mathbf{q}|}{\epsilon_0 \omega}\tilde{\sigma}},
\end{equation}
where $\mathbf{q}$ is the (in-plane) wave vector, $d$ is the distance between layers, $\sigma$ and $\tilde{\sigma}$ are the longitudinal conductivities of the alkali and hBN layers, respectively, and $\epsilon_0$ is the vacuum permittivity. Since the hBN bandgap \cite{cassabois2016hexagonal} is much larger than the plasmonic frequencies, we neglected the frequency dependence of the hBN conductivity and used values for the hBN dielectric constant from \cite{thygesen2017calculating}. We calculated the plasmon dispersion for five filling factors (\cref{figure: figure 2}(b)), $n$, with $n=0.5$ corresponding to the undoped system. Since the electronic bandwidth is 0.9 eV (\cref{figure: figure 1}(a)), any lossless plasmon at lowest order must exceed 1.08 eV, regardless of doping. However, there is also an upper limit, set by $E_{\text{c}}-E_{\text{Fermi}}-\omega_{\text{ph}}$. For the five doping values considered in this paper, only $n=0.1$ and $n=0.2$ yield plasmons with frequencies that lie within this first-order lossless range (\cref{figure: figure 2}(c)). For $n=0.1$, we find a maximum plasmon frequency of 1.16 eV, so all plasmons from 1.08 to 1.16 eV are lossless. For $n=0.2$, we find a maximum plasmon frequency of 1.44 eV, but only plasmons from 1.08 to 1.16 eV are lossless (plasmons above 1.16 eV may absorb a phonon and decay via an interband transition). 
\section{Plasmonic losses}
\begin{figure}
    \centering
\includegraphics{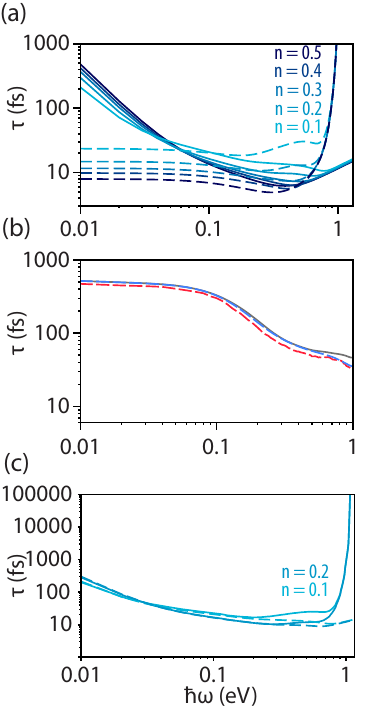}
    \caption{\textbf{Plasmonic decay rate due to electron-phonon interaction}. (a) Decay rates calculated through \cref{equation:final tau} (solid lines) and \cref{equation: wrong decay rate} (dashed lines) for the $\sqrt{7}\times\sqrt{7}$ sodium intercalated structure. \cref{equation: wrong decay rate} does not take into account the lifetime of intermediate states but explicitly accounts for the finite bandwidth and energy conservation at the one-phonon level, leading to a divergence at $W+\omega_\text{ph}$ and a comparatively small DC decay time. (b) Comparison of decay times as calculated through \cref{equation:final tau} (blue dashed line), \cref{equation: final tau 1} (grey line) and \cref{equation: wrong decay rate} (red dashed line) for graphene at 0.5 eV doping from the Dirac point (solid grey line). The agreement of the three decay time methods (for all frequencies) is due to the non-bandwidth-limited nature of the electronic structure at the Fermi level. (c) Decay times for the $\sqrt{7}\times\sqrt{7}$ sodium intercalated structure calculated through \cref{equation: final tau 1} (solid lines) compared to those calculated through \cref{equation:final tau} (dashed lines) for the two filling factors that yield first order lossless plasmons. \cref{equation:final tau} does not take into account the finite bandwidth while \cref{equation: final tau 1} takes into account both the finite bandwidth and higher order decay processes, the former being the reason for the discrepancy between the two methods at higher frequencies.}
    \label{figure: figure 3}
\end{figure}
We calculated the plasmonic decay rate using three different methods. We did this to isolate the effects of (1) the finite electronic bandwidth at the Fermi level and (2) higher-order loss processes. Finite bandwidth effects must be accounted to correctly capture the phase space of plasmonic losses at frequencies comparable to the bandwidth, and higher-order electron-phonon processes must be accounted to correctly describe plasmonic lifetimes at frequencies at which second order decay processes dominate over first order ones \cite{ghorashi2024highly}. The most general expression for the lifetime, which takes into account both effects explicitly is given by
\cite{allen2015electron}:
\small
\begin{multline}
    \tau^{-1}(\omega)=\int_{-\infty}^{\infty} \int_{-\infty}^\infty\Bigg[\Im\Bigg[\frac{f(\omega')-f(\omega+\omega')}{\omega+\omega'-\varepsilon+i\delta-\Sigma(\varepsilon, \omega+\omega')}\Bigg] \\\times\Im\Bigg[\frac{1}{\omega'-\varepsilon+i\delta-\Sigma(\varepsilon, \omega')} \Bigg] \frac{g(\varepsilon)}{g(\varepsilon_F)}\Bigg]\frac{\hbar\omega}{\pi}\text{d}\varepsilon \text{d}\omega',
    \label{equation: final tau 1}
\end{multline}\normalsize
where $\delta$ is a positive infinitesimal, $f(\omega)$ is the Fermi-Dirac distribution, $\varepsilon_F$ is the Fermi energy, and $g(\varepsilon)$ is the electronic density of states. \cref{equation: final tau 1} is strictly valid only above a material-dependent frequency, $\omega_0$ (which we determine shortly). In \cref{equation: final tau 1}, the bandwidth is explicitly taken into account through the density of states, $g(\varepsilon)$, and plasmonic decay processes up to second order in the electron-phonon interaction are taken into account through the electronic self-energy, $\Sigma(\varepsilon, \omega)$, calculated to lowest order in the electron-phonon coupling \cite{giustino2017electron}\footnote{In this work, we consider only the imaginary part of the self energy. In principle, the real part contributes to a change in the effective bandwidth.}. The Fermi surface averaged version of \cref{equation: final tau 1}, which does not take into account the finite bandwidth is given by \cite{allen2015electron}: 
\begin{equation}
\tau^{-1}(\omega) = \hbar\omega \int_{-\infty}^\infty \Im\Bigg[\frac{[f(\omega+\omega')-f(\omega')]}{\Sigma^*(\omega')-\Sigma(\omega+\omega')+\hbar\omega}\Bigg]\text{d}\omega',
\label{equation:final tau}
\end{equation}
where $\Sigma(\omega)\equiv \Sigma(\varepsilon_F, \omega)$ is the Fermi surface averaged electronic self-energy. As it is a Fermi surface averaged version of \cref{equation: final tau 1}, \cref{equation:final tau} is also only valid above the same material-dependent frequency, $\omega_0$. \cref{equation:final tau} corresponds to Fermi surface averaging of the correlation bubble in which both electronic propagators are corrected via a Dyson expansion (see SM). It is \emph{not} equivalent to Fermi-surface averaging of all one-phonon corrections to the correlation bubble (as in the approach used in \cite{sundararaman2020plasmonics}, see SM). In \cite{ghorashi2024highly}, we showed that the latter technique is equivalent to explicit calculation of the one-phonon decay rate via:  \small
\begin{multline}
            \tau^{-1}(\omega) = \frac{2\pi}{N_\mathbf{k} N_{\mathbf{k}'}\hbar^2\omega g(\varepsilon_F)}\sum_{\mathbf{k}, \mathbf{k}' j \pm}
        |g_{\mathbf{k},\mathbf{k}'}^{j}|^2\Big(N^{j, \mp}_{\mathbf{k}-\mathbf{k}'}f_\mathbf{k}- N_{\mathbf{k}-\mathbf{k}'}^{j, \pm}f_{\mathbf{k}'} \\ \pm f_\mathbf{k}f_{\mathbf{k}'} \Big) \delta(\epsilon_\mathbf{k}+\hbar\omega\pm \hbar\omega^j_{\mathbf{k}-\mathbf{k}'}-\epsilon_{\mathbf{k}'}) \Bigg(1-\frac{\mathbf{v}_\mathbf{k} \cdot \mathbf{v}_{\mathbf{k}'}}{|\mathbf{v}_\mathbf{k}||\mathbf{v}_{\mathbf{k}'}|}\Bigg), 
    \label{equation: wrong decay rate}
\end{multline}
\normalsize
where $N_\mathbf{k}$ is the number of sampled k-points, $g(\varepsilon_F)$ is the density of states at the Fermi level, 
$v_\mathbf{k}$, $v_{\mathbf{k}'}$ are the electronic velocities at wavevectors $\mathbf{k}$ and $\mathbf{k}'$, respectively, $\omega_\mathbf{q}^j$ are the phonon frequencies of branch $j$ at wavevector $\mathbf{q}$,
with corresponding $N^j_\mathbf{q}$ Bose occupation factors. In \cref{equation: wrong decay rate} we have defined the quantities $$N_\mathbf{q}^{j, \pm} \equiv \frac{1}{2} +N_\mathbf{q}^j \pm \frac{1}{2},$$ 
where the plus (minus) sign corresponds to phonon emission (absorption).
The sum in \cref{equation: wrong decay rate} includes all phonon bands, indexed by $j$ but includes only the isolated electronic band at the Fermi level. 

Unlike \cref{equation: final tau 1},  \cref{equation: wrong decay rate} does not take into account decay processes beyond first order in the electron-phonon interaction, but it does take into account the finite bandwidth. The advantage of \cref{equation: wrong decay rate}, however, is that it interpolates correctly between the DC and high-frequency limits \cite{ghorashi2024highly}. As a result, for a general, non-Drude metal, one may find the aforementioned frequency, $\omega_0$ at which \cref{equation: final tau 1} and \cref{equation: wrong decay rate} agree in order to determine the frequency region for which \cref{equation: final tau 1} is valid ($\omega>\omega_0$). From \cref{figure: figure 3}(a, c), we see that, in general, for our Moire systems, \cref{equation: final tau 1} is valid for frequencies above $\approx 0.1$ eV. 

For a system that is not bandwidth-limited (frequencies of interest, $\omega<<W$), and which is well-described by the Drude model, such as graphene, we find that the choice of decay time formula leads to negligible effects on the plasmonic lifetime (\cref{figure: figure 3}(b)). For a finite bandwidth, however, the difference may be significant (\cref{figure: figure 3}(a)). It should be noted that \cref{equation:final tau}  includes two-phonon processes, but since it does not take into account the finite bandwidth, it does not yield a correct result at frequencies comparable to the bandwidth. In \cref{figure: figure 3}(c), we show the decay time as calculated through \cref{equation: final tau 1}. As shown, \cref{equation: final tau 1} and \cref{equation:final tau} agree for frequencies up to $\approx 0.2$ eV. At frequencies in the lossless regime, we see from \cref{figure: figure 3}(c) that the decay time diverges (reaching $\approx 10^8$ fs for $\hbar\omega=1.16$ eV). At these frequencies, other scattering processes, e.g. two-plasmon scattering and impurity scattering will dominate (see next section and SM). 

\section{Plasmon to plasmon scattering}
In the absence of electron-phonon scattering, the only remaining intrinsic decay mechanism for a plasmon is through the electron-electron interaction. Electron-electron interaction mediated decay processes correspond to, for instance, a plasmon decaying into a lower frequency plasmon and an electron-hole pair. Such a process cannot be avoided by band structure engineering (since the maximum optical phonon frequency does not set a threshold for losslessness in this case). 

Plasmon-plasmon scattering has already been considered in the case of graphene \cite{jablan2015multiplasmon}. However, in \cite{jablan2015multiplasmon}, only the rate of absorption of two plasmons at the same frequency and at the same wavevector was considered. Here, we adopt a more general approach (in the SM we show that our approach is consistent with \cite{jablan2015multiplasmon}). We consider the effect of electron-electron scattering on the electron self-energy within the plasmon-pole approximation: 
\small
\begin{equation}
\begin{split}
        \Im\Sigma(\mathbf{k}, \omega')=-\sum_{\mathbf{q}}\frac{\pi e^2\omega_\mathbf{q}}{4N_\mathbf{k}\Omega\varepsilon_0|\mathbf{q}|}\times\\\Bigg[(1+N_\mathbf{q}-f_{\mathbf{k}+\mathbf{q}})\delta(\varepsilon_{\mathbf{k}+\mathbf{q}}-\hbar\omega'+\hbar\omega_\mathbf{q})+\\(f_{\mathbf{k}+\mathbf{q}}+N_\mathbf{q})\delta(\varepsilon_{\mathbf{k}+\mathbf{q}}-\hbar\omega'-\hbar\omega_\mathbf{q}) \Bigg],
        \end{split}
\end{equation}
\normalsize
where $\Omega$ is the unit cell area, $\omega_\mathbf{q}$ corresponds to the plasmon frequency at wavevecto $\mathbf{q}$ and the $N_\mathbf{q}$ now correspond to Bose occupation factors of the plasmons (not the phonons). We again consider only the isolated electronic band at the Fermi level. 

\begin{figure}
    \centering
\includegraphics{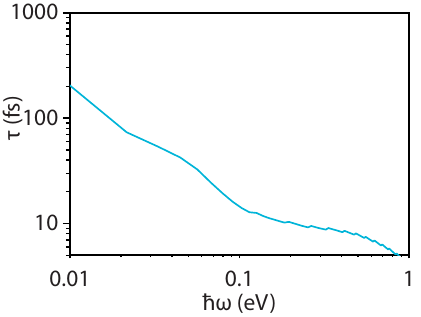}
    \caption{\textbf{Plasmon-plasmon scattering rate}. We show the rate of scattering of a plasmon of frequency $\omega$ with all other plasmons supported by the intercalated system (for $n=0.1$). The decay time takes into account both contributions from the electron-electron and electron-phonon interactions. At high frequencies, however, the electron-phonon interaction contributes negligibly to plasmon scattering, indicating that the high rate of decay at high frequencies is due to plasmon-plasmon scattering.} 
    \label{figure: figure 4}
\end{figure}

In \cref{figure: figure 4}, we evaluate \cref{equation: final tau 1} using the total self energy (both electron-electron and electron-phonon contributions) to show the effect of plasmon-plasmon scattering on the plasmon lifetime. As seen in \cref{figure: figure 4}, in the frequency regime where electron-phonon scattering is negligible $\hbar\omega>1$eV, the electron-electron interaction dominates. Plasmons at these high frequencies would be expected to decay at a rate of about $10^{14}s^{-1}$ to lower frequency plasmons.  

\section{Methods}
DFT calculations were implemented through JDFTx \cite{sundararaman2017jdftx} with the gga-PBE exchange-correlation functional \cite{perdew1996generalized}, norm-conserving pseudopotentials \cite{schlipf2015optimization}, Coulomb truncation \cite{sundararaman2013regularization} and Van-der-Waals D3 correction \cite{grimme2010consistent}. Plasmonic dispersions and losses were calculated using maximally localized Wannier orbitals \cite{marzari1997maximally, giustino2007electron} in conjunction with in-house code \cite{JJDFTX.jl}.
\section{Outlook}
We have shown that intercalated Moire heterostructures may be used to create plasmons that are lossless up to some order in the electron-phonon interaction. In addition, we have shown that the transport properties of these bandwidth-limited systems behave unconventionally compared to non-bandwidth-limited systems, such as graphene. In particular, we find that a correct description of the plasmonic lifetimes may only be found through careful consideration of higher-order processes as well as the finite electronic bandwidth. Importantly, we find that that sodium intercalation of Moire heterostructures of hBN can yield plasmons immune to first-order losses through the electron-phonon interaction, with electron-phonon associated quality factors many orders of magnitude higher than possible through known materials.

Importantly, however, we find that the electron-electron interaction makes these high frequency plasmons---which are immune to electron-phonon associated losses---decay into lower frequency plasmons. We stress that this prediction rests on the validity of the plasmon-pole approximation, the validity of which could be a subject for future work. 

\section{Acknowledgments}
The work is supported in part by the U. S. Army Research Office through the Institute for Soldier Nanotechnologies at MIT, under Collaborative Agreement Number W911NF-23-2-0121.








\bibliography{main}

\end{document}